# Effect of Ni, Fe Intercalation on the Superconducting Properties of ZrTe$_3$

C. S. Yadav and P. L. Paulose

*Tata Institute of Fundamental Research, Navynagar, Colaba, Mumbai-400005*
*Email: csyadav@tifr.res.in*

**Abstract.** We report the superconductivity at enhanced temperature of 5.2 K in the polycrystalline sample of ZrTe$_3$ and Ni intercalated ZrTe$_3$. ZrTe$_3$ is a Charge Density Wave (T = 63K) compound, which is known to superconduct only below 2K in single crystalline form. We discuss that the intergrain strains in the polycrystalline samples induces an intrinsic pressure and thus enhances the transition temperature. Fe intercalation of ZrTe$_3$ kills both the charge density wave and superconducting states, gives rise to the magnetic ordering in the compound.

The trichalcogenides of transition metal TX$_3$ (T = Ti, Zr, Hf, Nb, Nb, Ta; X = S, Se, Te) compounds are known for the Charge Density Wave (CDW) ground state at low temperatures. These compounds have a structure made of the infinite chains of TX6 extending along the b axis of the monoclinic chains. Among these NbSe$_3$ superconducts only under pressure and TiSe$_3$ become upon intercalation copper. The ZrTe$_3$ has also been found to show only filamentary superconductivity below 2 K temperature. Copper and Nickel intercalated single crystal compounds of ZrTe$_3$ are shown to exhibit superconducting (SC) properties at 3.8 K and 3.1 K respectively. In our recent studies we have found that the polycrystalline sample of ZrTe$_3$, and its Cu and Ag intercalated analogues, prepared at higher temperature, show the superconductivity at elevated temperature of 5.2 K. The superconducting state was found to coexist with the CDW state, and intercalation of Cu, Ag slightly weakens the CDW but raises it to higher temperatures. In the present studied we report the effect of intercalation of magnetic atoms Ni and Fe on the superconducting and CDW states of the ZrTe$_3$ polycrystalline compounds. We have also studied the effect of intercalation on the magnetic properties of the compound.

Polycrystalline samples of ZrTe$_3$, Ni$_{0.05}$ZrTe$_3$, and Fe$_{0.05}$ZrTe$_3$ were prepared from the chemical reaction of the high purity elements Zr, Te, Ni, and Fe, inside the evacuated (10$^{-6}$ mbar pressure) quartz tubes at 975$^0$C for 48 hours. The obtained charge was found to be reacted completely. It was further grounded before magnetic measurement. For resistivity and heat capacity measurements, the compounds were pelletized at 10 ton pressure. The X Ray-diffraction patterns of the compounds were using Philips X'pert PRO-Diffractometer in Brag Brentano geometry with a slow scan rate. QD-MPMS (SQUID) and PPMs were used for the magnetic and electrical measurements respectively.

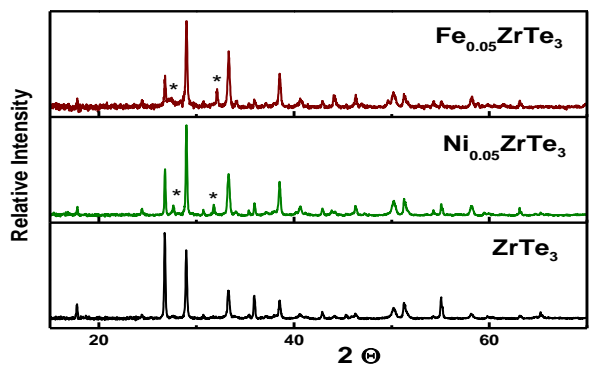

**FIGURE 1.** XRD pattern of ZrTe$_3$, Ni$_{0.05}$ZrTe$_3$, and Fe$_{0.05}$ZrTe$_3$ compounds. The peaks marked with star (*) shows the enhanced reflection of planes.

The XRD pattern of ZrTe$_3$, Ni$_{0.05}$ZrTe$_3$, and Fe$_{0.05}$ZrTe$_3$ compounds is shown in the figure 1. The patterns were fitted with GSAS Rietveld program within the monoclinic (Space group: P2$_1$m), with the Lattice constants a = 5.863Å, b = 3.924 Å, c = 10.093Å, and β = 97.740 for ZrTe3. It is difficult to notice any appreciable change in lattice parameters for the lightly intercalated (~ 5%) compounds using out XRD data. The relative change in the intensity of Bragg reflections is a signature of intercalation. SEM (EDAX) also confirms the presence of Ni and Fe in the ZrTe$_3$.

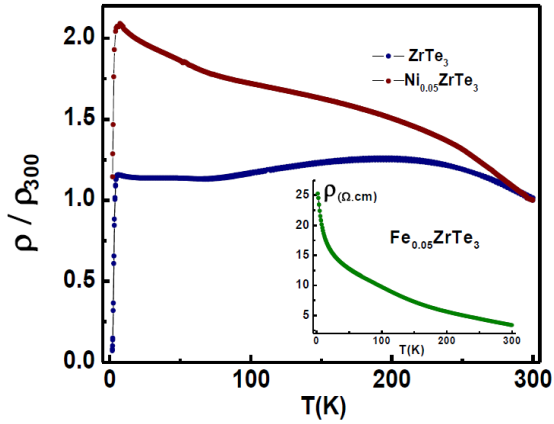

**FIGURE 2.** Normalized Resistivity of $ZrTe_3$ and $Ni_{0.05}ZrTe_3$ and the Inset shows the resistivity of $Fe_{0.05}ZrTe_3$.

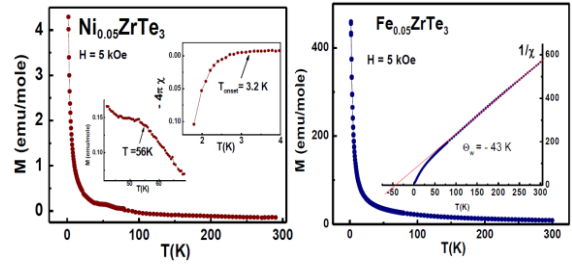

**FIGURE 3.** M vs T for $Ni_{0.05}ZrTe_3$, and $Fe_{0.05}ZrTe_3$. The inset (a) shows Superconducting and CDW transition $Ni_{0.05}ZrTe_3$ (b) Curie fit to $1/\chi$ vs T showing $\Theta w = -43K$ for $Fe_{0.05}ZrTe_3$.

appreciable change in lattice parameters for the lightly intercalated (~ 5%) compounds using out XRD data The relative change in the intensity of Bragg reflections is a signature of intercalation. SEM (EDAX) also confirms the presence of Ni and Fe in the $ZrTe_3$.

The Electrical resistivity of the compounds of the $ZrTe_3$ and $Ni_{0.05}ZrTe_3$ shows superconductivity below 5.2 K. This temperature is much higher than the reported below 2 K for $ZrTe_3$. The transition width for $Ni_{0.05}ZrTe_3$ is wider than in for $ZrTe_3$. Similar to the observation in $NbSe_3$, this increase in $T_c$ in polycrystalline compounds can be attributed to the internal strain and pressure between the small crystallites of the compounds. The upper critical field $H_{c2}(0)$ value of the compounds are also found to be low as 1.3 T. The room temperature resistivities are 10mΩ-cm, 75 mΩ-cm, and 3 Ω-cm for $ZrTe_3$ and $Ni_{0.05}ZrTe_3$ and $Fe_{0.05}ZrTe_3$ respectively. As seen from the figure 2, in the normal state, resistivity of compounds has increased on intercalation of Ni. However the Fe intercalated compound doesn't superconduct and remain insulating down to 1.8 K (inset of figure 2). The insulating behavior of the compound cannot be fit using either band model, or the variable hopping model and seems to be arising from the weak localization of charge carrier due to various scattering process in these polycrystalline samples. The CDW transition in $ZrTe_3$ and $Ni_{0.05}ZrTe_3$ is clearly visible in resistivity, where the depletion of the charge carriers on account of the opening of band gap at CDW transition leads to the rise in resistivity of the compounds.

The figure 3 show the magnetization data for $Ni_{0.05}ZrTe_3$ and $Fe_{0.05}ZrTe_3$ taken at H = 5 kOe. The top inset of the fig 3(a) shows the superconducting transition at 3.2 K for $Ni_{0.05}ZrTe_3$. Similar to $ZrTe_3$ compound (ref), superconducting fraction in this compound is also low (~ 12%), and the broad transition is a signature of distribution of $T_c$ within the compound, which is possible when the uneven internal strain in the crystallite grains lead to different Tc. The low temperature Curie tail lead to the $\mu_{eff} = 0.09\mu_B$. The lower inset of the figure shows the CDW anomaly 56K. The lower of CDW is in conformity with the electrical resistivity results where the resistance increases faster upon after this temperature. Contrary to $Ni_{0.05}ZrTe_3$, the $Fe_{0.05}ZrTe_3$ is highly magnetic in nature (figure 3(b)), and no superconductivity is observed. The inset of the figure show the Curie-Weiss fit to the $1/\chi$ vs T and gives the $T_N = -43K$. The value of magnetic moment $\mu_{eff} = 1.30\mu_B$. The superconductivity is reported 5.2 K in polycrystalline $ZrTe_3$. The intercalation of Ni does not affect SC $T_c$ and the CDW phase is retained albeit at lower temperatures. The SC is filamentary in nature with a distribution of $T_c$ within the material. Since single crystal of the compounds shows SC at lower temperature, the rise in $T_c$ for these compounds can be understood in terms of the enhanced intergrain strains in polycrystalline samples that induce an intrinsic pressure which is akin to the effect of external pressure on SC $T_c$. The intercalation of magnetic atom Fe kills CDW as well as SC states, and gives rise to Antiferromagnetic ordering. The Fe intercalation seem to bring a change in the electronic structure, and thus changes the topology of the Fermi surface (FS) thus killing the both FS driven phenomenon of CDW and SC.

We acknowledge Manish Ghagh and Prasad Mundye for the help in sample preparation and measurement.